\documentclass[lnbip]{svmultln}

\usepackage{makeidx}

\usepackage{todonotes}

\usepackage{graphicx}
\usepackage{epsfig}
\usepackage{latexsym}
\usepackage{colortbl}
\usepackage{times}
\usepackage{helvet}
\usepackage{courier}
\usepackage{caption}
\usepackage{subfigure}
\usepackage{float}
\usepackage{xcolor}
\usepackage{pslatex}
\usepackage{eso-pic}
\usepackage[bottom]{footmisc}
\usepackage{url}
\usepackage{etex}
\usepackage{xspace}
\usepackage{alltt}
\usepackage{pgfplots}
\usepackage{tikz}
\usepackage{comment}
\usepackage{adjustbox}
\usepackage{wrapfig,fancyvrb}

\usepackage{blindtext,graphicx}
\usepackage[absolute]{textpos}
\begin{document}
\begin{textblock}{16}(3,1)
\noindent\Large Preprint from Proc. of 1st Int. Workshop on Cognitive Business Process Management (CBPM 2017)
\end{textblock}

\begin{sloppypar}

\newcommand {\myi} {\emph{(i)}~}
\newcommand {\myii} {\emph{(ii)}~}
\newcommand {\myiii} {\emph{(iii)}~}
\newcommand {\myiv} {\emph{(iv)}~}
\newcommand {\myv} {\emph{(v)}~}
\newcommand {\myvi} {\emph{(vi)}~}

\newcommand{\DD}{\mbox{$\cal D$}}                     
\newcommand{\PP}{\mbox{$\cal P$}}                     
\newcommand{\CC}{\mbox{$\cal C$}}                     
\newcommand{\RR}{\mbox{$\cal R$}}                     


\newcommand{\exogDom}{\fontFluents{Exog}}
\newcommand{\noOp}{\fontActions{noOp}}

\newcommand{\inp}{i}
\newcommand{\out}{o}
\newcommand{\capa}{c}
\newcommand{\id}{\fontFluents{id}}
\newcommand{\srv}{\fontFluents{srv}}
\newcommand{\SRV}{\fontFluents{SRV}}

\newcommand{\tinst}[2]{#1\!:\!#2}

\newcommand{\ExpOut}{\fontFluents{ExpOut}}
\newcommand{\InputArg}{\fontFluents{Input}}
\newcommand{\fAdapted}{\fontFluents{MustAlign}}
\newcommand{\aUnsetAdapted}{\fontActions{UnsetMustAlign}}
\newcommand{\aSetAdapted}{\fontActions{SetMustAlign}}

\def\prparallel{\mathrel{\rangle\!\rangle}}
\def\supparallel{\mathord{|\!|}}
\newcommand{\conc}{\mbox{$\parallel$}}
\newcommand{\pconc}{\mbox{$\prparallel$}}
\newcommand{\mnt}{\mbox{$mnt$}}

\newcommand{\set}[1]{\{#1\}}

\newcommand{\Pre}{\fontFluents{Pre}}
\newcommand{\Eff}{\fontFluents{Eff}}
\newcommand{\Par}{\fontFluents{Par}}


\newcommand{\hilight}[1] {\colorbox{yellow}{#1}}

\newcommand{\propername}[1]{\textsf{#1}\xspace}

\newcommand{\DDSPM}{\DD_{\smartpm}}
\newcommand{\ExogProg}{\delta_{\text{\textit{exog}}}}
\newcommand{\smartpm}{\propername{SmartPM}}
\newcommand{\smartpmSIM}{\propername{SmartPM Simulator}}
\newcommand{\indigolog}{\propername{IndiGolog}}
\newcommand{\ConGolog}{\propername{ConGolog}}
\newcommand{\Golog}{\propername{Golog}}
\newcommand{\golog}{\propername{Golog}}
\newcommand{\LPGtd}{\propername{LPG-td}}
\newcommand{\PDDL}{\propername{PDDL}}

\newcommand{\citeby}[1]{\citeauthor{#1}~(\citeyear{#1})}

\newcommand {\smartpmsystem} {\propername{SmartPM System}}
\newcommand {\smartpmML} {\propername{SmartPM Modeling Language}}
\newcommand {\smartpmDT} {\propername{SmartPM Definition Tool}}
\newcommand {\smartML} {\propername{SmartML}}
\newcommand {\swiprolog} {\textsf{SWI-Prolog}}
\newcommand {\SWIProlog} {\textsf{SWI-Prolog}}
\newcommand{\Prol}[1]{\texttt{\mbox{#1}}}
\newcommand{\kTrue}{\texttt{kTrue}}
\newcommand{\mTrue}{\texttt{mTrue}}

\newcommand{\arrow}[1]{\,\hbox{$\stackrel{\small \mbox{#1}}{\rightarrow}$}\,}
\newcommand{\arrowstar}{\,\hbox{$\stackrel{\mbox{$\ast$}}{\rightarrow}$}\,}
\newcommand{\isdef}{\ \hbox{$\stackrel{\mbox{\tiny def}}{=}$}\ }

\newcommand{\fontFluents}[1]{\mbox{\textit{#1}}}
\newcommand{\fontActions}[1]{\mbox{\textsc{#1}}}

\newcommand{\Poss}{\fontFluents{Poss}}
\newcommand{\Trans}{\fontFluents{Trans}}
\newcommand{\Final}{\fontFluents{Final}}
\newcommand{\Do}{\fontFluents{do}}

\newcommand{\aUnlock}{\fontFluents{unlock}}
\newcommand{\aOpenDoor}{\fontActions{open}}
\newcommand{\aCloseDoor}{\fontActions{close}}
\newcommand{\fDoorOpen}{\fontFluents{DoorOpen}}
\newcommand{\fFloor}{\fontFluents{Floor}}
\newcommand{\fLocked}{\fontFluents{Locked}}
\newcommand{\fColor}{\fontFluents{Color}}
\newcommand{\fExpOut}{\fontFluents{ExOut}}
\newcommand{\fFree}{\fontFluents{Free}}
\newcommand{\fReserved}{\fontFluents{Reserved}}
\newcommand{\fRunning}{\fontFluents{Running}}
\newcommand{\fAssigned}{\fontFluents{Assigned}}
\newcommand{\fReady}{\fontFluents{Ready}}
\newcommand{\fIdentifier}{\fontFluents{Identifier}}
\newcommand{\fHolding}{\fontFluents{Holding}}
\newcommand{\fCompleted}{\fontFluents{Completed}}

\def\prparallel{\mathrel{\rangle\!\rangle}}
\newcommand{\mif}{\mbox{\bf if}}
\newcommand{\mforeach}{\mbox{\bf for each}}
\newcommand{\mforeachin}{\mbox{\bf in}}
\newcommand{\mwhile}{\mbox{\bf while}}
\newcommand{\mreturn}{\mbox{\bf return}}
\newcommand{\mthen}{\mbox{\bf then}}
\newcommand{\melse}{\mbox{\bf else}}
\newcommand{\mdo}{\mbox{\bf do}}
\newcommand{\mnoOp}{\mbox{\bf noOp}}
\newcommand{\mproc}{\mbox{\bf proc}}
\newcommand{\mProc}{\text{\bf Proc}\xspace}
\newcommand{\mend}{\mbox{\bf end}}
\newcommand{\mendproc}{\mbox{\bf endProc}}
\newcommand{\mendif}{\mbox{\bf endIf}}
\newcommand{\mendwhile}{\mbox{\bf endWhile}}
\newcommand{\mendforeach}{\mbox{\bf endFor}}
\newcommand{\mendfor}{\mbox{\bf endFor}}
\newcommand{\mfor}{\mbox{\bf for}}
\def\supparallel{\mathord{|\!|}}
\newcommand{\tuple}[1]{\ensuremath{\langle #1 \rangle}}     
\newcommand{\search}{\mbox{$\Sigma$}}
\newcommand{\nil}{\mbox{\it nil}}
\newcommand{\false}{\mathtt{false}}
\newcommand{\False}{\mathtt{false}}
\newcommand{\true}{\mathtt{true}}
\newcommand{\True}{\mathtt{True}}
\newcommand{\xdo}{\mbox{\it do}}

\newcommand{\recovered}{\fontFluents{Recovered}}
\newcommand{\Monitor}{\fontFluents{Monitor}}
\newcommand{\Relevant}{\fontFluents{Misaligned}}
\newcommand{\fRelevant}{\fontFluents{Relevant}}
\newcommand{\SameConfig}{\fontFluents{SameConfig}}
\newcommand{\SameState}{\fontFluents{SameState}}
\newcommand{\Linear}{\fontFluents{Linear}}
\newcommand{\Recovery}{\fontFluents{Recovery}}
\newcommand{\fRealityChanged}{\fontFluents{RealityChanged}}

\newcommand{\fLocation}{\fontFluents{Location\_type}}
\newcommand{\fLoc}{\fontFluents{Location}}
\newcommand{\fObj}{\fontFluents{CeramicElement}}

\newcommand{\fState}{\fontFluents{Status\_type}}
\newcommand{\fInteger}{\fontFluents{Integer\_type}}
\newcommand{\fBoolean}{\fontFluents{Boolean\_type}}

\newcommand{\fService}{\fontFluents{Service}}
\newcommand{\fActor}{\fontFluents{Actor}}
\newcommand{\fRobot}{\fontFluents{Robot}}
\newcommand{\fSensor}{\fontFluents{Sensor}}
\newcommand{\fMovingRobot}{\fontFluents{MovingRobot}}
\newcommand{\workitem}{\fontFluents{Workitem}}
\newcommand{\fTask}{\fontFluents{Task}}
\newcommand{\ftask}{\fontFluents{task}}
\newcommand{\fCapability}{\fontFluents{Capability}}
\newcommand{\fProvides}{\fontFluents{Provides}}
\newcommand{\fRequires}{\fontFluents{Requires}}
\newcommand{\fNeigh}{\fontFluents{Neigh}}
\newcommand{\fCovered}{\fontFluents{Covered}}

\newcommand{\fFresh}{\fontFluents{FreshId}}

\newcommand{\fCapable}{\fontFluents{Capable}}
\newcommand{\fAvailable}{\fontFluents{Available}}
\newcommand{\fListElem}{\fontFluents{ListElem}}
\newcommand{\fVar}{\fontFluents{Var}}
\newcommand{\fX}{\fontFluents{X}}
\newcommand{\fF}{\fontFluents{F}}
\newcommand{\fY}{\fontFluents{Y}}
\newcommand{\fEnabled}{\fontFluents{Reserved}}
\newcommand{\DomainOfX}{\fontFluents{DomainOfX}}

\newcommand{\catalano}{\fontActions{Catalano}}

\newcommand{\aAssign}{\fontActions{assign}}
\newcommand{\aStart}{\fontActions{start}}
\newcommand{\aAck}{\fontActions{ackCompl}}
\newcommand{\aRelease}{\fontActions{release}}
\newcommand{\aReady}{\fontActions{readyToStart}}
\newcommand{\aFinish}{\fontActions{finished}\xspace}
\newcommand{\aGo}{\fontActions{go}}
\newcommand{\aConvey}{\fontActions{convey}}
\newcommand{\aGlaze}{\fontActions{glaze}}
\newcommand{\aMould}{\fontActions{mould}}
\newcommand{\aCheckQuality}{\fontActions{checkquality}\xspace}
\newcommand{\aGoUp}{\fontActions{up}}
\newcommand{\aGoDown}{\fontActions{down}}
\newcommand{\aMove}{\fontActions{move}}
\newcommand{\aTakePhoto}{\fontActions{takephoto}}
\newcommand{\aEvacuate}{\fontActions{evacuate}}
\newcommand{\aRemoveDebris}{\fontActions{removedebris}}
\newcommand{\aExtinguishFire}{\fontActions{extinguishfire}}
\newcommand{\aUpdateStatus}{\fontActions{updatestatus}}
\newcommand{\aChargeBattery}{\fontActions{chargeBattery}}
\newcommand{\aPushed}{\fontActions{pushed}}
\newcommand{\aFixTemperature}{\fontActions{fixtemp}}
\newcommand{\aHighTemperature}{\fontActions{hightemp}}
\newcommand{\aDeposit}{\fontActions{deposit}}
\newcommand{\aGetId}{\fontActions{getid}}

\newcommand{\expectedReality}{\fontActions{\textbf{Exp}}\xspace}
\newcommand{\physicalReality}{\fontActions{\textbf{Phy}}\xspace}

\newcommand{\started}{\fontActions{started}}
\newcommand{\assigned}{\fontActions{assigned}}
\newcommand{\notassigned}{\fontActions{not assigned}}

\newcommand{\aAdaptStart}{\fontActions{adaptStart}}
\newcommand{\aAdaptFinish}{\fontActions{adaptFinish}}
\newcommand{\aResetReality}{\fontActions{resetReality}}

\newcommand{\fAt}{\fontFluents{At}}
\newcommand{\fAtRobot}{\fontFluents{AtRobot}}
\newcommand{\fEvacuated}{\fontFluents{Evacuated}}
\newcommand{\fStatus}{\fontFluents{Status}}
\newcommand{\fPhotoTaken}{\fontFluents{PhotoTaken}}

\newcommand{\fMoveStep}{\fontFluents{MoveStep}}
\newcommand{\fDebrisStep}{\fontFluents{DebrisStep}}
\newcommand{\fBatteryRecharging}{\fontFluents{BatteryRecharging}}
\newcommand{\fGeneralBattery}{\fontFluents{GeneralBattery}}

\newcommand{\send}{\fontActions{send}}
\newcommand{\alignRealities}{\fontActions{align}}
\newcommand{\abortTask}{\fontActions{abort}}
\newcommand{\receive}{\fontActions{receive}}

\newcommand{\isConnected}{\fontFluents{Connected}}
\newcommand{\atLeastOne}{\fontFluents{atLeastOne}}
\newcommand{\atLeastAnotherOne}{\fontFluents{atLeastAnotherOne}}
\newcommand{\isRobotConnected}{\fontFluents{isRobotConnected}}

\newcommand{\fEnoughBattery}{\fontFluents{EnoughBattery}}
\newcommand{\fBatteryLevel}{\fontFluents{BatteryLevel}}

\newcommand{\fPhotoOK}{\fontFluents{PhotoOK}}
\newcommand{\fRescueOK}{\fontFluents{RescueOK}}
\newcommand{\fSurveyOK}{\fontFluents{SurveyOK}}
\newcommand{\aPhoto}{\fontActions{photo}}
\newcommand{\aSurvey}{\fontActions{survey}}
\newcommand{\aRescue}{\fontActions{rescue}}

\newcommand{\photoLost}{\fontActions{photolost}}
\newcommand{\rockSlide}{\fontActions{rockslide}}
\newcommand{\fireRisk}{\fontActions{firerisk}}

\newcommand{\fAdapting}{\fontFluents{Adapting}}

\newcommand {\ps} {\emph{Planning Service}}
\newcommand {\planlet} {\textsc{Planlet}}
\newcommand {\planlets} {\textsc{Planlets}}

\newcommand{\proc}[1]{{\textsf{#1}}}
\newcommand{\fExog}{\fontFluents{Exogenous}}
\newcommand{\fExogAction}{\fontFluents{Exog}}
\newcommand{\fAffectedExog}{\fontFluents{InteractExogTask}}
\newcommand{\aResetExog}{\fontActions{resetExog}}
\newcommand{\fIsExog}{\fontFluents{ExogAction}}
\newcommand{\fIsFinished}{\fontFluents{Finished}}
\newcommand{\finish}{\fontActions{finish}}
\newcommand{\wait}{\fontActions{wait}}
\newcommand{\halt}{\fontActions{halt}}

%
%
%
%
%
%
%

\newcommand{\finishEnvironment}{\hspace*{\fill}\qed}


\newcommand{\FP}{FP}
\newcommand{\SC}{SC}

\pagestyle{plain}

\title{Cognitive Business Process Management\\ for Adaptive Cyber-Physical Processes}
\titlerunning{Cognitive BPM for Adaptive CPPs}  
%
\toctitle{Cognitive Business Process Management for Adaptive Cyber-Physical Processes}

\author{Andrea Marrella \and Massimo Mecella}
\authorrunning{A. Marrella and M. Mecella} 
%
%
\institute{
Sapienza Universit\`a di Roma\\ Dipartimento di Ingegneria Informatica, Automatica e Gestionale\\ via Ariosto 25, 00185 Roma, Italy\\
\email{[lastname]@diag.uniroma1.it}
}

\maketitle              

\begin{abstract}
In the era of Big Data and Internet-of-Things (IoT), all real-world environments are gradually becoming cyber-physical (e.g., emergency management, healthcare, smart manufacturing, etc.), with the presence of connected devices and embedded ICT systems (e.g., smartphones, sensors, actuators) producing huge amounts of data and events that influence the enactment of the Cyber Physical Processes (CPPs) enacted in such environments.
A Process Management System (PMS) employed for executing CPPs is required to automatically adapt its running processes to anomalous situations and exogenous events by minimising any human intervention at run-time.
In this paper, we tackle this issue by introducing an approach and an adaptive Cognitive PMS that combines process execution monitoring, unanticipated exception detection and automated resolution strategies leveraging on well-established action-based formalisms in Artificial Intelligence, which allow to interpret the ever-changing knowledge of cyber-physical environments and to adapt CPPs by preserving their base structure.
\end{abstract}


\section{Introduction}
\label{sec:introduction}

In the last years, we have witnessed the emergence of new computing paradigms, such as Industry 4.0\footnote{cf. H. Kagermann, W. Wahlster and J. Helbig: Recommendations for implementing the strategic initiative Industrie 4.0: Final report of the Industrie 4.0 Working Group, 2013, Frankfurt, \url{http://www.acatech.de/fileadmin/user_upload/Baumstruktur_nach_Website/Acatech/root/de/Material_fuer_Sonderseiten/Industrie_4.0/Final_report__Industrie_4.0_accessible.pdf}
}, Health 2.0 (e.g., cf. \cite{cossu2012improving}) and mobile-based emergency management \cite{humayoun2009designing,humayoun2009workpad}, in which the interplay of Internet-of-Things (IoT) devices, i.e., devices attached to the Internet, cloud computing, Software-as-a-Service (SaaS), and Business Process Management (BPM) create the so-called \emph{cyber-physical environments} and give rise to the concept of \emph{Cyber-Physical Systems} (CPSs).
The role of CPSs is to monitor the \emph{physical processes} enacted in cyber-physical environments, create a virtual copy of the physical world and make decentralized decisions, by introducing automated and intelligent support of workers in their increasingly complex work~\cite{Seiger2014}.

A relevant aspect in these environments lies in the fundamental role played by the processes orchestrating the different actors (software, humans, robots, etc.) involved in the CPS. We refer to these processes as \emph{cyber-physical processes} (CPPs), whose enactment is influenced by user decision making and coupled with contextual data and knowledge production coming from the cyber-physical environment. According to~\cite{Hull@BPM2016}, \emph{Cognitive Process Management Systems} (CPMSs) are the key technology for supporting CPPs. A PMS is said to be \emph{cognitive} when it involves additional processing constructs that are at a semantic level higher than those of conventional PMSs. These constructs are called \emph{cognitive BPM constructs} and include data-driven activities, goals, and plans~\cite{Hull@BPM2016}. Their usage can open opportunities for new levels of automation for CPPs, such as - for example - \emph{the automated synthesis of adaptation strategies at run-time exploiting solely the process knowledge and its expected evolution}.

During the enactment of CPPs, variations or divergence from structured reference models are common due to exceptional circumstances arising (e.g., autonomous user decisions, exogenous events, or contextual changes), thus requiring the ability to properly \emph{adapt} the process behavior. \emph{Process adaptation} can be seen as the ability of a process to react to exceptional circumstances (that may or may not be foreseen) and to adapt/modify its structure accordingly. %
Exceptions can be either \emph{anticipated} or \emph{unanticipated}. An anticipated exception can be planned at design-time and incorporated into the process model, i.e., a (human) process designer can provide an \emph{exception handler} that is invoked during run-time to cope with the exception. Conversely, \emph{unanticipated exceptions} refer to situations, unplanned at design-time, that may emerge at run-time and can be detected only during the execution of a process instance, when a mismatch between the computerized version of the process and the corresponding real-world process occurs. To cope with those exceptions, a PMS is required to allow \emph{ad-hoc process changes} for adapting running process instances in a context-dependent way.

The fact is that, in cyber-physical environments, the number of possible anticipated exceptions is often too large, and traditional manual implementation of exception handlers at design-time is not feasible for the process designer, who has to anticipate all potential problems and ways to overcome them in advance~\cite{ReichertBook2012}. Furthermore, anticipated exceptions cover only partially relevant situations, as in such scenarios many unanticipated exceptional circumstances may arise during the process instance execution. Therefore, the process designer often lacks the needed knowledge to model all the possible exceptions at the outset, or this knowledge can become obsolete as process instances are executed and evolve, by making useless her/his initial effort.

To tackle this issue, in this paper we summarize the main ideas discussed in~\cite{AIComm} and introduce our work on \smartpm\footnote{The reader interested to the very technical details may refer to ~\cite{KR2014,DBLP:journals/tist/MarrellaMS17,AIComm}.}, a CPMS able to \emph{automatically adapt CPPs at run-time} when \emph{unanticipated exceptions} occur, thus requiring no specification of recovery policies at design-time. The general idea builds on the dualism between an \emph{expected reality} and a \emph{physical reality}: process execution steps and exogenous events have an impact on the physical reality and any deviation from the expected reality results in a mismatch to be removed to allow process progression.

To that end, we have resorted to three popular \emph{action-based formalisms} and \emph{technologies} from the field of Knowledge Representation and Reasoning (KR\&R): situation calculus~\cite{ReiterBook}, \indigolog~\cite{Indigolog:2009}, and automated planning~\cite{TraversoBook2004,GeffnerBonet:BOOK13-PLANNING}.
We used the situation calculus logical formalism to model the underlying domain in which processes are to be executed, including the description of available tasks, contextual properties, tasks' preconditions and effects, and the initial state.
On top of such model, we used the \indigolog high-level agent programming language for the specification of the structure and control flow of processes. Importantly, we customized \indigolog to monitor the online execution of processes and detect potential mismatches between the model and the actual execution. If an exception invalidates the enactment of the processes being executed, an external state-of-the-art classical planner is invoked to synthesise a recovery procedure to adapt the faulty process instance.

The choice of adopting action-based formalisms from the KR\&R field is motivated by their ability to provide the right \emph{cognitive} level needed when dealing with dynamic situations in which data (values) play a relevant role in system enactment and automated reasoning over the system progress.
In the field of BPM, many other formalisms (in particular Petri Nets-based and process algebras) have been successfully adopted for process management, but all of them are somehow based on synthesis techniques of the control-flow, when considering their automated reasoning capabilities. This implies the level of abstraction over dealing with data and dynamic situations is fairly ``raw'', when compared with KR\&R methods in which automated reasoning over data values and situations is much more developed~\cite{reichgelt1991knowledge,Brachman:2004,ReiterBook}.
The choice of KR\&R technologies allowed us to develop a principled, clean and simple-to-manage framework for process adaptation based on relevant data manipulated by the process, without compromising efficiency and effectiveness of the proposed solution.

The rest of the paper is organized as follow. Section \ref{sec:background} introduces conceptual architecture for CPMSs that manage CPPs. Such an architecture is then instantiated in the \smartpm approach and system outlined in Section \ref{sec:architecture}. Finally Section \ref{sec:discussion} concludes the paper by discussing our approach in the larger context and presenting possible future evolutions.

\section{A Conceptual Architecture for Managing CPPs}
\label{sec:background}

CPSs are having widespread applicability and proven impact in multiple areas, like aerospace, automotive, traffic management, healthcare, manufacturing, emergency management~\cite{rajkumar2010cyber}. According to~\cite{lee2008cyber}, any physical environment which contains computing-enabled devices can be considered as a cyber-physical environment.

The trend of managing CPPs, i.e., processes enacted in cyber-physical environments, has been fueled by two main factors. On the one hand, the recent development of powerful mobile computing devices providing wireless communication capabilities have become useful to support mobile workers to execute tasks in such dynamic settings. On the other hand, the increased availability of sensors disseminated in the world has lead to the possibility to monitor in detail the evolution of several real-world objects of interest. \emph{The knowledge extracted from such objects allows to depict the contingencies and the context in which processes are carried out, by providing a fine-grained monitoring, mining, and decision support for them.}

We devise in the following a conceptual architecture 
to concretely build  an adaptive CPMS in cyber-physical environments. The management of a CPP requires additional challenges to be considered if compared with a traditional ``static'' business process.
On the one hand, there is the need of representing explicitly real-world objects and technical aspects like device capability constraints, sensors range, actors and robots mobility, etc. On the other hand, since cyber-physical environments are intrinsically ``dynamic'', a CPMS providing real-time monitoring and automated adaptation features during process execution is required.

To this end, the role of the data perspective becomes fundamental. Data, including information processed by process tasks as well as contextual information, are the main driver for triggering process adaptation, as focusing on the control flow perspective only - as traditional PMSs do - would be insufficient. In fact, in a cyber-physical environment, a CPP is genuinely knowledge and data centric: its control flow must be coupled with contextual data and knowledge production and process progression may be influenced by user decision making. This means that traditional imperative models have to be extended and complemented with the introduction of specific \emph{cognitive constructs} such as \emph{data-driven activities} and \emph{declarative elements} (e.g., tasks preconditions and effects) which enable a precise description of data elements and their relations, so as to go beyond simple process variables, and allow establishing a link between the control flow and the data perspective.

\begin{figure}
	\centering
	\includegraphics[width=0.65\columnwidth]{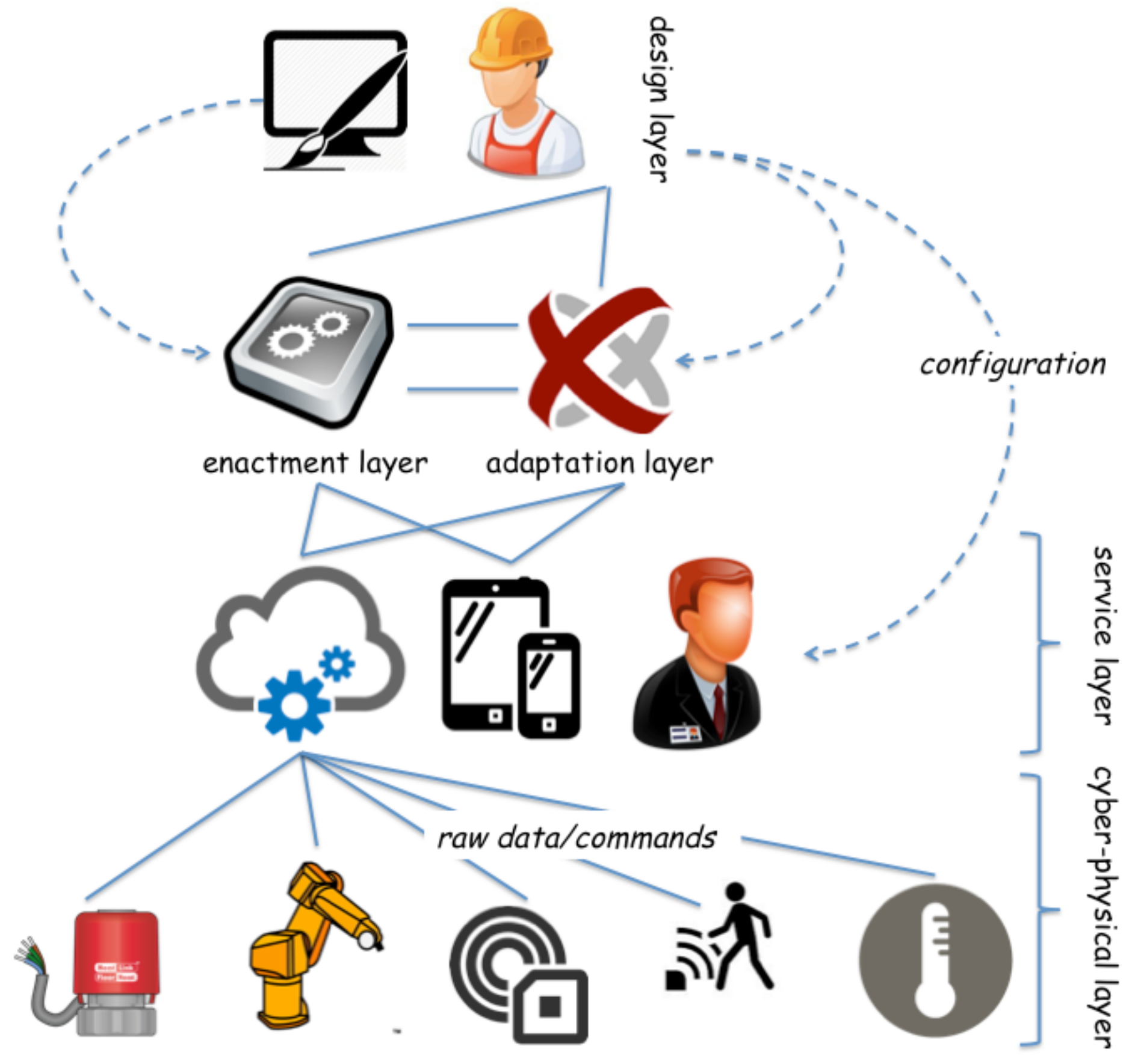}
	\caption{A conceptual architecture for CPPs.}
	\label{fig:conceptual_architecture}
\end{figure}

Starting from the above considerations, coupled with the experience gained in the area and lessons learned from several projects involving CPSs, we have devised a conceptual architecture to build a CPMS for the management of CPPs, which supports the so-called \emph{Plan-Act-Learn} cycle for cognitively-enabled processes~\cite{Hull@BPM2016}.
As shown in Fig.~\ref{fig:conceptual_architecture}, we identified 5 main architectural layers that we present in a bottom-up fashion.

The \emph{\textbf{cyber-physical layer}} consists mainly of two classes of physical components: \myi sensors (such as GPS receivers, RFID chips, 3D scanners, cameras, etc.) that collect data from the physical environment by monitoring real-world objects and \myii actuators (robotic arms, 3D printers, electric pistons, etc.), whose effects affect the state of the physical environment. The cyber-physical layer is also in charge of providing a physical-to-digital interface, which is used to transform \emph{raw} data collected by the sensors into machine-readable events, and to convert \emph{high-level} commands sent by the upper layers into \emph{raw} instructions readable by the actuators. The cyber-physical layer does not provide any intelligent mechanism neither to clean, analyse or correlate data, nor to compose high-level commands into more complex ones; such tasks are in charge of the uppers layer.

On top of the cyber-physical layer lies the \emph{\textbf{service layer}}, which contains the set of services offered by the real-world entities (software components, robots, agents, humans, etc.) to perform specific process tasks.
In the service layer, available data can be aggregated and correlated, and high-level commands can be orchestrated to provide higher abstractions to the upper layers.
For example, a smartphone equipped with an application allowing to sense the position and the posture of a user is at this layer, as it collects the raw GPS, accellerometer and motion sensor data and correlates them to provide discrete and meaningful information.

On top of the service layer, there are two further layers interacting with each other. The \emph{\textbf{enactment layer}} is in charge of \myi enacting complex processes by deciding which tasks are enabled for execution, \myii orchestrating the different available services to perform those tasks and \myiii providing an execution monitor to detect the anomalous situations that can possibly prevent the correct execution of process instances. The execution monitor is responsible for deciding if process adaptation is required. If this is the case, the \emph{\textbf{adaptation layer}} will provide the required algorithms to \myi reason on the available process tasks and contextual data and to \myii find a recovery procedure for adapting the process instance under consideration, i.e., to re-align the process to its expected behaviour. Once a recovery procedure has been synthesized, it is passed back to the enactment layer for being executed.

Finally, the \emph{\textbf{design layer}} provides a GUI-based tool to define new process specifications. A process designer must be allowed not only to build the process control flow, but also to explicitly formalize the data reflecting the contextual knowledge of the cyber-physical environment under study. It is important to underline that data formalization must be performed without any knowledge of the internal working of the physical components that collect/affect data in the cyber-physical layer.
To link tasks to contextual data, the GUI-based tool must go beyond the classical ``task model'' as known in the literature, by allowing the process designer to explicitly state what data may constrain a task execution or may be affected after a task completion. Finally, besides specifying the process, configuration files should also be produced to properly configure the enactment, the services and the sensors/actuators in the bottom layers.

\section{The \smartpm Approach and System}
\label{sec:architecture}

\smartpm (Smart Process Management) is an approach and an adaptive CPMS implementing a set of techniques that enable to automatically adapt process instances at run-time in the presence of unanticipated exceptions, without requiring an explicit definition of handlers/policies to recover from tasks failures and exogenous events.
\smartpm adopts a \emph{layered} \emph{service-based} approach to process management, i.e., \emph{tasks are executed by services}, such as software applications, humans, robots, etc. Each task can be thus seen as a single step consuming input data and producing output data.

\begin{figure}[t]
\centering{
 \includegraphics[width=0.8\textwidth]{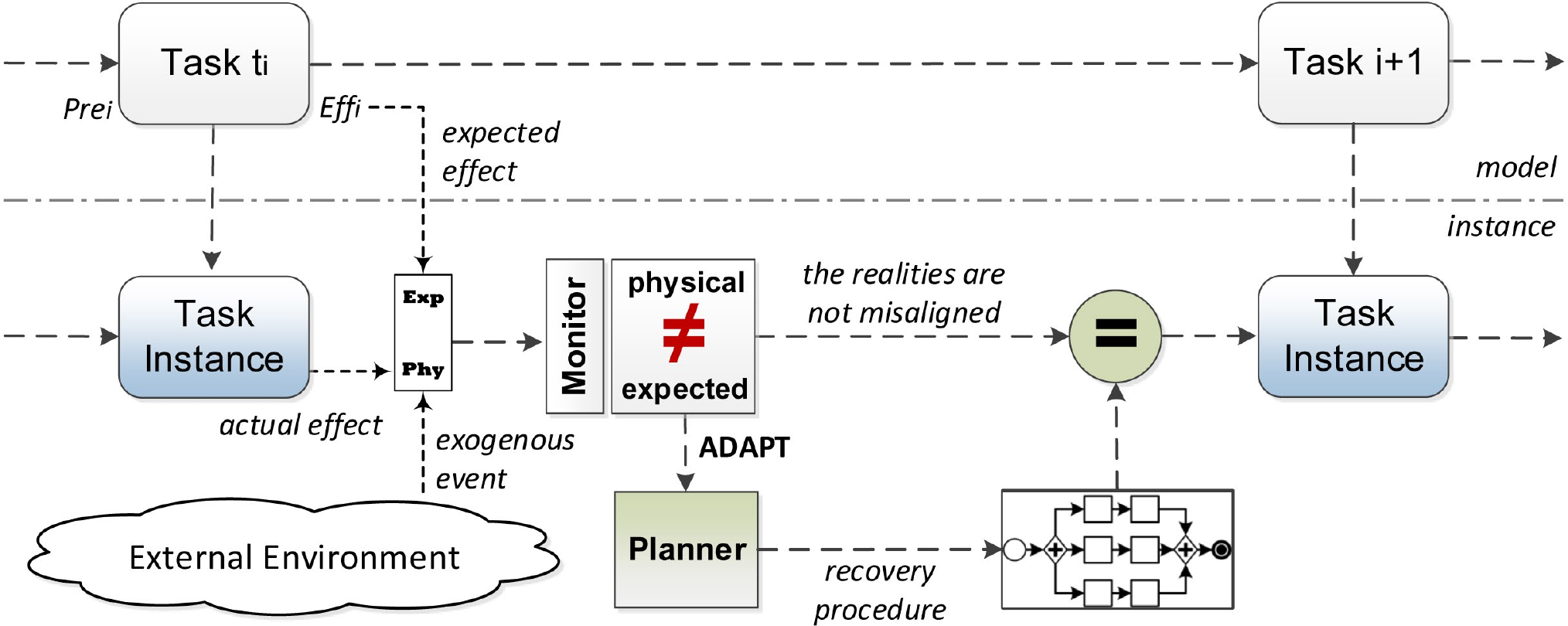}
 }
 \caption{An overview of the \smartpm approach.}
 \label{fig:smartpm_approach}
\end{figure}

To monitor and deal with exceptions, the \smartpm approach leverages on \cite{GiacomoRS98}'s technique of adaptation from the field of agent-oriented programming, by specializing it to our CPP setting (see Fig.~\ref{fig:smartpm_approach}).
We consider adaptation as \emph{reducing the gap} between the \emph{expected reality} \expectedReality, the (idealized) model of reality used by the CPMS to reason, and the \emph{physical reality} \physicalReality, the real world with the actual conditions and outcomes.
While \physicalReality records what is \emph{concretely} happening in the real environment during a process execution, \expectedReality reflects what it is \emph{expected} to happen in the environment. Process execution steps and exogenous events have an impact on \physicalReality and any deviation from \expectedReality results in a mismatch to be removed to allow process progression.
At this point, a state-of-the-art automated planner is invoked to synthesise a recovery procedure that adapts the faulty process instance by removing the gap between the two realities.

To realize the above approach, the implementation of \smartpm covers the modeling, execution and monitoring stages of the CPP life-cycle. 
To that end, the architecture of \smartpm relies on five architectural layers.

The \emph{\textbf{design layer}} provides a graphical editor developed in Java that assists the process designer in the definition of the process model at design-time. Process knowledge is represented as a \emph{domain theory} that includes all the contextual information of the domain of concern, such as the people/services that may be involved in performing the process, the tasks, the data and so forth.
Data are represented through some \emph{atomic terms} that range over a set of \emph{data objects}, which depict entities of interest (e.g., capabilities, services, etc.), while atomic terms can be used to express properties of domain objects (and relations over objects). \emph{Tasks} are collected in a repository and are described in terms of \emph{preconditions} - defined over atomic terms - and \emph{effects}, which establish their expected outcomes. Finally, a process designer can specify which \emph{exogenous events} may be caught at run-time and which atomic terms will be modified after their occurrence.
Once a valid domain theory is ready, the process designer uses the graphical editor to define the process control flow through the standard BPMN notation among a set of tasks selected from the tasks repository.
%

The \emph{\textbf{enactment layer}} is in charge of managing the process execution. First of all, the domain theory specification and the BPMN process are automatically translated into situation calculus~\cite{ReiterBook} and \indigolog~\cite{Indigolog:2009} readable formats. Situation calculus is used for providing a declarative specification of the domain of interest (i.e., available tasks, contextual properties, tasks preconditions and effects, what is known about the initial state). Then, an \emph{executable model} is obtained in the form of an \indigolog program to be executed through an \indigolog engine.
To that end, we customized an existing \indigolog engine\footnote{\texttt{http://sourceforge.net/projects/indigolog/}} to \myi build a physical/expected reality by taking the initial context from the external environment; \myii manage the process routing; \myiii collect exogenous events from the external environment; \myiv monitor contextual data to identify changes or events which may affect process execution. Once a task is ready for being executed, the \indigolog engine assigns it to a proper process participant (that could be a software, a human actor, a robot, etc.) that provides all the required capabilities for task execution. 

The \textbf{\emph{service layer}} acts as a middleware between process participants, the enactment layer and the cyber-physical layer.
Specifically, in the service layer, process participants interact with the engine through a \emph{Task Handler}, an interactive GUI-based software application realized for Android devices that supports the visualization/execution of assigned tasks by selecting an appropriate outcome.
Possibly such an Android application can exploit sensors and actuators (e.g., an Arduino board connected through Bluetooth, as currently realized in our implementation), thus effectively offering services over the \emph{cyber-physical layer}.
Every step of the task life cycle - ranging from the assignment to the release of a task - requires an interaction between the \indigolog engine and the task handlers. 
The communication between the \indigolog engine and the task handlers is mediated by the \emph{Communicator Manager} component (which is essentially a web server) and established using the Google Cloud Messaging service.
%


To enable the automated synthesis of a recovery procedure, the \emph{\textbf{adaptation layer}} relies on the capabilities provided by a planner component (the LPG-td planner~\cite{LPG}), which assumes the availability of a classical planning problem, i.e., an initial state and a goal to be achieved, and of a planning domain definition that includes the actions to be composed to achieve the goal, the domain predicates and data types. Specifically, if process adaptation is required, we translate \myi the domain theory defined at design-time into a planning domain, \myii the physical reality into the initial state of the planning problem and \myiii the expected reality into the goal state of the planning problem. The planning domain and problem are the input for the planner component.
If the planner is able to synthesize a recovery procedure $\delta_a$, the \emph{Synchronization} component combines $\delta'$ (which is the remaining part of the faulty process instance $\delta$ still to be executed), with the recovery plan $\delta_a$, builds an adapted process $\delta'' = (\delta_a;\delta')$ and converts it into an executable \indigolog program so that it can be enacted by the \indigolog engine.
Otherwise, if no plan exists for the current planning problem, the control passes back to the process designer, who can try to manually adapt the process instance.


The \emph{\textbf{cyber-physical layer}} is tightly coupled with the physical components available in the domain of interest.
Since the \indigolog engine can only work with defined discrete values, while data gathered from physical sensors have naturally continuous values, the system provides several web tools that allow process designers to associate some of the data objects defined in the domain theory with the continuous data values collected from the environment. For example, we developed several web tools to associate the data collected from sensors (GPS, temperature, noise level, etc.) to discrete values. We provided a concrete example of a location web tool that allows process designers to mark areas of interest from a real map and associate them to discrete locations. The mapping rules generated are then saved into the Communication Manager and retrieved at run-time to allow the matching of the continuous data values collected by the specific sensor into discrete data objects.

\section{Concluding Remarks}
\label{sec:discussion}

We are at the beginning of a profound transformation of BPM due to advances in AI and Cognitive Computing \cite{Hull@BPM2016}. Cognitive systems offer computational capabilities typically based on large amount of data, which provide cognition power that augment and scale human expertise. The aim of the emergent field of cognitive BPM  is to offer the computational capability of a cognitive system to provide analytical support for processes over structured and unstructured information sources. The target is to provide proactivity and self-adaptation of the running processes against the evolving conditions of the application domains in which they are enacted.

In this direction, our paper summarizes the most interesting results reported in~\cite{AIComm}, which have been devoted to the realization of a general approach, a concrete framework and a CPMS implementation, called \smartpm, for automated adaptation of CPPs. Our purpose was to demonstrate that the combination of procedural and imperative models with cognitive BPM constructs such as data-driven activities and declarative elements, along with the exploitation of techniques from the field of AI such as situation calculus, \indigolog and classical planning, can increase the ability of existing PMSs of supporting and adapting CPPs in case of unanticipated exceptions.

Existing approaches dealing with unanticipated exceptions typically rely on the involvement of process participants at run-time, so that authorized users are allowed to manually perform structural process model adaptation and ad-hoc changes at the instance level. However, CPPs demand a more flexible approach recognizing the fact that in real-world environments process models quickly become outdated and hence require closer interweaving of modeling and execution. To this end, the adaptation mechanism provided by \smartpm is based on execution monitoring for detecting failures and context changes at run-time, without requiring to predefine any specific adaptation policy or exception handler at design-time (as most of the current approaches do).

From a general perspective, our planning-based automated exception handling approach should be considered as complementary with respect to existing techniques, acting as a ``bridge'' between approaches dealing with anticipated exceptions and approaches dealing with unanticipated exceptions. When an exception is detected, the run-time engine may first check the availability of a predefined exception handler, and if no handler was defined it can rely on an automated synthesis of the recovery process.
In the case that our planning-based approach fails in synthesizing a suitable handler (or an handler is generated but its execution does not solve the exception), other adaptation techniques need to be used. For example, if the running process provides a well-defined intended goal associated to its execution, we could resort to the van Beest's work~\cite{vanBeest2014} and do planning from first-principle to achieve such a goal. Conversely, if no intended goal is associated to the process, a human participant can be involved, leaving her/him the task of manually adapting the process instance.
Future work will include an extension of our approach to ``stress'' the assumptions imposed by the usage of classical planning techniques for the synthesis of the recovery procedure, which frame the scope of applicability of the approach for addressing more expressive problems, including incomplete information, preferences and multiple task effects.

The current implementation of \smartpm is developed to be effectively used by process designers and practitioners.\footnote{See: \url{http://www.dis.uniroma1.it/~smartpm}} Users define processes in the well-known BPMN language, enriched with semantic annotations for expressing properties of tasks, which allow our interpreter to derive the \indigolog program representing the process. Interfaces with human actors (such as specific graphical user applications in Java) and software services (through Web service technologies) allow the core system to be effectively used for enacting processes. Although the need to explicitly model process execution context and annotate tasks with preconditions and effects may require some extra modeling effort at design-time (also considering that traditional process modeling efforts are often mainly directed to the sole control flow perspective), the overhead is compensated at run-time by the possibility of automating exception handling procedures. While, in general, such modeling effort may seem significant, in practice it is comparable to the effort needed to encode the adaptation logic using alternative methodologies like happens, for example, in rule-based approaches.

%
%
\section*{Acknowledgments} This work is partly supported by the projects Social Museum and Smart Tourism (CTN01\_00034\_23154), NEPTIS (PON03PE\_00214\_3), RoMA (SCN\_00064), and by the Sapienza project ``Data-aware Adaptation of Knowledge-intensive Processes in Cyber-Physical Domains through Action-based Languages''.

\bibliographystyle{splncs}
\bibliography{references,additional-refs}

\end{sloppypar}
\end{document}